\documentclass[nohyper,twoside]{JHEP3}

\usepackage{graphicx}
\usepackage{bm}
\usepackage{amsmath}

\newcommand{\be}{\begin{equation}}
\newcommand{\ee}{\end{equation}}
\newcommand{\bea}{\begin{eqnarray}}
\newcommand{\eea}{\end{eqnarray}}
\newcommand{\bdm}{\begin{displaymath}}
\newcommand{\edm}{\end{displaymath}}
\newcommand{\beas}{\begin{eqnarray*}}
\newcommand{\eeas}{\end{eqnarray*}}

\newcommand{\mD}{\mathcal{D}}

\newcommand{\av}[1]{\left< #1\right>}
\newcommand{\dom}[1]{#1_{\mathcal{D}}}
\newcommand{\intdom}[1]{\frac{1}{\dom{V}}\int_{\mathcal{D}} #1\sqrt{h}d^3\mathbf{x}}
\newcommand{\Hd}{\frac{\dom{\dot{a}}}{\dom{a}}}
\newcommand{\Vd}{V_{\mathcal{D}}}
\newcommand{\Rd}{\mathcal{R}_{\mathcal{D}}}
\newcommand{\Pd}{\mathcal{P}_{\mathcal{D}}}

\newcommand{\Qd}{\mathcal{Q}_{\mathcal{D}}}

\newcommand{\Fd}{\mathcal{F}_{\mathcal{D}}}
\newcommand{\Td}{\mathcal{T}_{\mathcal{D}}}
\newcommand{\Sd}{\mathcal{S}_{\mathcal{D}}}
\newcommand{\Ld}{\mathcal{L}_{\mathcal{D}}}

\newcommand{\bkr}{\overline{\rho}}
\newcommand{\bkp}{\overline{p}}

\newcommand{\pd}{\partial}
\newcommand{\hub}{\frac{\dot{a}}{a}}

\preprint{HD-THEP-09-19}

\title{Accelerating the Universe with Gravitational Waves}

\author{Iain A. Brown$^1$, Lily Schrempp$^1$ and Kishore Ananda$^2$
\\ \email{I.Brown@thphys.uni-heidelberg.de}, \email{L.Schrempp@thphys.uni-heidelberg.de}, \email{Kishore.Ananda@gmail.com}
\\ 1: Institut f\"ur Theoretische Physik, Philosophenweg 16, 69120 Heidelberg, Germany
\\ 2: Department of Mathematics and Applied Mathematics, University of Cape Town, Rondebosch 7701, Cape Town, South Africa}

\received{\today}

\abstract{
Inflation generically produces primordial gravitational waves with a red spectral tilt. In this paper we calculate the backreaction produced by these gravitational waves on the expansion of the universe. We find that in radiation domination the backreaction acts as a relativistic fluid, while in matter domination a small dark energy emerges with an equation of state $w_\mathrm{eff}\approx -8/9$.
}

\begin{document}

\section{Introduction}
The discovery of an apparent universal acceleration in the late 1990s \cite{RiessEtAl98,PerlmutterEtAl98} lead to a multitude of theoretical explanations. Amongst the most appealing of these, given that they involve no new physics, are those based on the inhomogeneous structure of spacetime. Spatial inhomogeneities can influence cosmological models in two main ways: they alter the null geodesics and so affect photon propagation and the luminosity distance $d_L(z)$, and due to the nonlinearity of the Einstein tensor they cause the evolution of the average cosmology to deviate from the Robertson-Walker behaviour. In the language of \cite{Kolb09}, these are dubbed ``weak cosmological backreaction'' and ``strong cosmological backreaction'' respectively, although it might be noted that the term ``backreaction'' typically refers to the impact of perturbations on their background. The weak backreaction -- deviations of the null geodesics from Robertson-Walker predictions -- is generally studied with reference to LeMa\^itre-Tolman-Bondi (LTB) models (see for example \cite{Alnes06,Marra07,Biswas07,Vanderveld06,Enqvist08,Vanderveld08,Krasinski09,Celerier09} and their references for a small sample of the recent literature) and we will not consider it further here. The so-called ``strong backreaction'' can be formulated in general coordinate systems \cite{Larena09,Brown09-2} and has a long history \cite{Shirokov63,Marochnik80,Marochnik80-2,Ellis84,Futamase89,Kasai93,Mukhanov96}. The discovery of the universal acceleration lead to an increase in interest in the field \cite{Boersma97,Russ97,Buchert99,Buchert01,Wetterich03,Rasanen03}. Current research focuses on extending the formalism (see for example \cite{Larena09,Brown09-2,Korzynski09}) and on quantifying the impact in particular systems \cite{Brown09,Paranjape09,Clarkson09}. There has been much recent work in the field, and we point the interested reader to Buchert's 2007 review \cite{Buchert07} and the references found therein and in \cite{Brown09} for further references.

In this paper we focus on the backreaction arising in perturbed Friedmann-LeMa\^itre-Robertson-Walker (FLRW) universes, which serve as good models for the real universe across a wide range of redshifts. The perturbations to this model can be separated into scalar, vector and tensor components based on how they transform on spatial three-surfaces. Of the three classes, the scalars have been widely considered since the earliest studies \cite{Shirokov63,Marochnik80,Marochnik80-2} of cosmological averaging, since scalars are the dominant perturbation and develop strong inhomogeneities under structure formation. Marochnik \cite{Marochnik80,Marochnik80-2} also considered the vector and tensor modes, which have not received much attention since. Tensor modes decay as radiation and one would na\"ively expect the backreaction they induce at the present day to be many orders of magnitude beneath that from the scalars -- conversely, it is conceivable that, deep in radiation domination the tensor backreaction could be equivalent to or greater than that from scalars. The consideration of tensor modes is complicated by our lack of a simple covariant averaging procedure -- the so-called Buchert approach, which employs a 3+1 spatial slicing and averages only scalars, cannot meaningfully be applied to tensor quantities. In principle, the covariant averaging approach of Zalaletdinov \cite{Zalaletdinov96} can be applied to tensor modes, although attention with this approach has focused on solutions exhibiting a high symmetry \cite{Coley05,Hoogen09} or on scalar modes (e.g. \cite{Paranjape07}). Recent suggested averaging schemes by Behrend \cite{BehrendThesis08}, Gromes \cite{Gromes09,Gromes09-2}, Korzynski \cite{Korzynski09} and Coley \cite{Coley09} should also in principle be applicable to tensor modes. In this paper we will employ the standard 3+1 scalar averaging approach, focusing on the scalar quantities characterising the gravitational waves.

Scalar backreaction produces multiple correction terms. We employ the terminology and approach of \cite{Behrend08} (hereafter Paper I), \cite{Brown09} (hereafter Paper II) and \cite{Brown09-2} (Paper III). The corrections induced by the inhomogeneities are separated into six components: $\Rd$, a correction to the Friedmann equation that arises from the average spatial curvature; the kinematical backreaction $\Qd$, which appears in both the Friedmann and Raychaudhuri equations; the dynamical backreaction $\Pd$, which appears in the Raychaudhuri equation; and $\Td$, $\Sd$ and $\Ld$ which correct the fluid and cosmological constant properties. Paper II considered the scalar backreaction for a range of redshifts $z\in(0,10^4)$, finding that $\Rd$ is at least an order of magnitude greater than $\Pd$, while $\Qd$ is insignificant. This holds both in matter and in radiation domination. The effective equation of state of the backreaction evolves from $w_\mathrm{eff}\approx 1/3$ in radiation domination towards $w_\mathrm{eff}\approx 1/60$ at the present day\footnote{It should be noted that this figure is evaluated in Paper II assuming that averages of second-order perturbations of the form $\av{\phi_{(2)}}$ vanish. This will not in general be true; see for example \cite{Clarkson09}. In Paper III it was argued that this remains approximately true for systems where $\av{\phi_{(1)}^2}\approx\av{\phi_{(2)}}$.}.  In radiation domination, despite the effective equation of state, the effective energy density of the backreaction is approximately constant, while in matter domination it grows with the scale factor $a$. For a few recent studies focusing on scalar modes see \cite{Li08,Behrend08,Paranjape08,Brown09,Clarkson09}; the results are broadly consistent with one another when evaluated on the same scales, although the approaches are significantly different.

In the absence of active sources such as primordial magnetic fields or cosmic defects, there are two main mechanisms by which cosmological gravitational waves are produced: at linear order, directly from inflation, and at second-order, sourced by the linear scalar modes. While one might na\"ively expect the latter to be subdominant, they can instead at some periods and on some scales be of equivalent magnitude to the linear modes \cite{Ananda06,Baumann07,Assadullahi09}. This could be particularly significant around the epoch of matter/radiation equality \cite{Baumann07}. The analysis in an averaging context of such modes will be non-trivial due to the complicated form of their time-dependant power spectrum. In contrast, the statistics of the linear modes are imprinted at inflation and do not evolve. Furthermore, the equations of motion allow for analytic solutions, perhaps leading to closed forms for the backreaction they induce. We choose, therefore, to focus here only on the linear modes, giving us a good impression of the type of impacts gravitational radiation can have, while being amenable to analytic study. Modifications from the second-order modes would then be expected to be of at most the same order of magnitude, and we leave them to future study.

As compared to the scalar case, one would perhaps expect the curvature correction $\Rd$ induced by inflationary gravitational waves to remain dominant. However, there is no reason to believe that the tensor backreaction terms $\Qd$ and $\Pd$ should remain subdominant or evolve in the same manner as those from the scalars; unlike the scalar perturbations, tensor modes are unaffected by structure formation. However, we would expect significant differences between the nature of the backreaction that arises in radiation domination and that which arises in matter domination; in radiation domination a first-order tensor mode evolves with conformal time $\eta$ as $h\sim j_0(k\eta)$, while in matter domination it does so as $h\sim h_1j_1(k\eta)/(k\eta)+h_2y_1(k\eta)/(k\eta)$. This rapid decay in matter domination will induce differences between the evolutions of the correction terms $\Rd$, $\Qd$ and $\Pd$, leading perhaps to distinctive differences between the effective energy density and pressure in the very early and in the recent universe. Since the tensor modes are not affected by structure formation, unlike the scalars the effective energy density of the backreaction in matter domination should not be expected to grow rapidly; while this suggests that it should be small in the present epoch, it conversely suggests there could be a period in the past when the tensor backreaction dominates the scalar. The two main aims of this paper are to consider these problems: firstly, to determine the impact of the tensor modes deep in radiation domination and whether it is ever plausible that it is comparable to the scalar modes; and secondly, to investigate the behaviour of the tensor backreaction when the tensor modes decay in matter domination, and the nature of it today.

In \S\ref{Sec:Formalism} we set up the averaging formalism which we will employ, and in \S\ref{Sec:TensorPerturbations} we apply it to an FLRW universe containing scalar and tensor perturbations. \S\ref{Sec:Results} presents our findings and we end with a brief discussion in \S\ref{Sec:Discussion}.

\section{Formalism}
\label{Sec:Formalism}
Let spacetime be foliated on large scales with a family of 3-surfaces, choosing a coordinate system with a vanishing shift vector and the line element
\be
  ds^2=-\alpha^2dt^2+h_{ij}dx^idx^j .
\ee
The surfaces have the normal 4-vector
\be
  n^\mu=(1/\alpha,\mathbf{0}), \quad n_\mu=(-\alpha,\mathbf{0})
\ee
with the projection tensor onto the surface
\be
  h_{\mu\nu}=g_{\mu\nu}+n_\mu n_\nu .
\ee
The energy density and isotropic stress on this surface are
\be
  \varrho=n^\mu n^\nu T_{\mu\nu}, \quad
  S=h^{ij}S_{ij}=h^{ij}h^\mu_ih^\nu_jT_{\mu\nu}=h^{\mu\nu}T_{\mu\nu}
\ee
and are related to the rest-frame energy density $\rho$ and isotropic pressure $p$ of a fluid with 4-velocity $u^\mu$ by
\be
  \varrho=\rho+\left(n^\mu n^\nu-u^\mu u^\nu\right)T_{\mu\nu}, \quad
  S=3p+\left(n^\mu n^\nu-u^\mu u^\nu\right)T_{\mu\nu} .
\ee
Following for example Paper III we define a volume, average and an average volume expansion in a domain $\mathcal{D}$ lying on the 3-surface by
\be
  \Vd=\int_\mathcal{D}\sqrt{h}d^3\mathbf{x}, \quad \av{A}=\intdom{A}, \quad 3\Hd=\frac{\dot{V}_\mathcal{D}}{\Vd}
\ee
where $h$ is the determinant of the spatial 3-metric. Then volume-averaged Friedmann and Raychaudhuri equations can be derived from the Hamilton constraint and evolution of the extrinsic curvature respectively,
\bea
\label{BuchertHub}
  \left(\Hd\right)^2&=&\frac{8\pi G}{3}\av{\alpha^2\rho}+\frac{1}{3}\av{\alpha^2}\Lambda-\frac{1}{6}\left(\Rd+\Qd-6\Fd\right), \\
\label{BuchertRay}
  \frac{\dom{\ddot{a}}}{\dom{a}}&=&-\frac{4\pi G}{3}\av{\alpha^2\left(\rho+3p\right)}+\frac{1}{3}\av{\alpha^2}\Lambda+\frac{1}{3}\left(\Pd+\Qd-3\Fd\right) .
\eea
Here the curvature correction, kinematical backreaction, dynamical backreaction, and tilt correction are
\bea
  \Rd&=&\av{\alpha^2\mathcal{R}}, \\
  \Qd&=&\av{\alpha^2\left(K^2-K^i_jK^j_i\right)}-\frac{2}{3}\av{\alpha K}^2, \\
  \Pd&=&\av{\alpha\mD^i\mD_i\alpha}-\av{\dot{\alpha}K}, \\
  \Fd&=&\frac{8\pi G}{3}\av{\alpha^2\left(n^\mu n^\nu-u^\mu u^\nu\right)T_{\mu\nu}} .
\eea
Here $K_{ij}$ is the extrinsic curvature and $\mathcal{R}$ the Ricci scalar on the 3-surface. These expressions are valid for any system for which the 3+1 split holds.

\section{Tensor Perturbations}
\label{Sec:TensorPerturbations}
In Paper III the general backreaction expressions in a spatially flat Robertson-Walker universe perturbed to second-order with line element
\be
 ds^2=a^2(\eta)\left(-\left(1+2\phi\right)d\eta^2+2B_ad\eta dx^a+\left(\delta_{ij}+2C_{ij}\right)dx^idx^j\right)
\ee 
were presented. We wish to consider a universe containing scalar and tensor perturbations, neglecting the vector modes which, at linear order, decay rapidly. It is important to note that, while the tensor modes at linear order are intrinsically gauge-invariant, gauge issues remain a problem when considering their backreaction for two chief reasons. Firstly, the spatial volume is dependant upon $\sqrt{h}$ and, secondly, the tensor modes will couple to gradients of the scalar modes. We must therefore specify a gauge in which to work. Although a strong argument can be made for working in flat gauge \cite{Brown09-2}, we work for simplicity in Poisson gauge, which takes the line element
\be
  ds^2=a^2(\eta)\left(-\left(1+2\phi\right)d\eta^2+\left((1-2\psi)\delta_{ij}+h_{ij}^T\right)dx^idx^j\right)
\ee
where the tensor component of the 3-metric is gauge-invariant and transverse-traceless,
\be
  \delta^{ij}h^T_{ij}=\pd^ih_{ij}^T=0 .
\ee
The same caveats highlighted in Papers II and III concerning the averaging of a perturbation theory also apply here.

The curvature correction and the kinematical and dynamical backreactions are given by equations (40, 42, 43) of Paper III specialised to the case of a vanishing shift. Taking
\be
  C_{ij}=-\psi\delta_{ij}+\frac{1}{2}h_{ij}^T
\ee
and separating the corrections to the background into their scalar parts and tensor contributions, the curvature correction becomes
\be
\label{TensorR}
  \Rd=\Rd^{(S)}+\av{h^{ij}_T\pd^k\pd_kh_{ij}^T+4h^T_{ij}\pd^i\pd^j\psi
   +\frac{3}{4}\left(\pd^ih^{jk}_T\right)\left(\pd_ih_{jk}^T\right)
   -\frac{1}{2}\left(\pd^ih^{jk}_T\right)\left(\pd_jh_{ik}^T\right)} .
\ee
The scalar contribution $\Rd^{(S)}$ is given in equation (82) of Paper III and does not concern us here. The kinematical and dynamical backreactions can likewise be written
\be
\label{TensorBacks}
  \Qd=\Qd^{(S)}-\frac{1}{4}\av{\dot{h}_{ij}^T\dot{h}^{ij}_T}, \quad
  \Pd=\Pd^{(S)}-\av{h_{ij}^T\left(\pd^i\pd^j\phi+\frac{1}{2}\hub\dot{h}^{ij}_T\right)} .
\ee
Their scalar components are given in equations (80) and (81) respectively of Paper III. For a perfect-fluid source the tilt correction $\Fd$ is unchanged by the presence of gravitational waves, as are the subsequent modifications $\Td$, $\Sd$ and $\Ld$ which additionally correct for the non-unity lapse. The errors introduced by the coupling of the gravitational waves and the anisotropic stresses of the relativistic fluids is expected to be subdominant. The tensor corrections are intrinsically quadratic in that the leading-order backreaction arises from first-order squared terms and can thus be calculated good to second order in perturbation theory using linear perturbations and without complications from the metric determinant $\sqrt{h}$. From (\ref{BuchertHub}, \ref{BuchertRay}), the effective energy density and pressure of the backreaction can be identified as
\bea
  \frac{8\pi G}{3}a^2\bkr_\mathrm{eff}&=&\frac{8\pi G}{3}a^2\bkr_\mathrm{eff}^{(S)}
   -\frac{1}{6}\left(\Qd^{(T)}+\Rd^{(T)}\right), \\
  \frac{8\pi G}{3}a^2\bkp_\mathrm{eff}&=&\frac{8\pi G}{3}a^2\bkp_\mathrm{eff}^{(S)}
   +\frac{1}{18}\left(\Rd^{(T)}-3\Qd^{(T)}-4\Pd^{(T)}\right) .
\eea

In the remainder of this paper we will employ the normalised variables
\bea
\label{EnergyPressureE}
  \Omega_{\mathrm{eff}}^{(T)}=\frac{8\pi G}{3\mathcal{H}^2}a^2\bkr_\mathrm{eff}^{(T)}=-\frac{1}{6(\dot{a}/a)^2}\left(\Rd^{(T)}+\Qd^{(T)}\right), \\
\label{EnergyPressureP}
  \Pi_{\mathrm{eff}}^{(T)}=\frac{8\pi G}{3\mathcal{H}^2}a^2\bkp_\mathrm{eff}^{(T)}=\frac{\Rd^{(T)}-3\Qd^{(T)}-4\Pd^{(T)}}{18(\dot{a}/a)^2}
\eea
and
\be
\label{wRay}
  w_\mathrm{eff}^{(T)}=\frac{\Omega_\mathrm{eff}^{(T)}}{\Pi_\mathrm{eff}^{(T)}}, \quad
  \left(\frac{\Delta R}{R}\right)^{(T)}=\frac{1}{3\left|\left(\dot{a}/a\right)^\cdot\right|}\left(\Pd^{(T)}+\Qd^{(T)}\right)
\ee
to characterise the contribution of the tensor modes to the background. These correspond to the effective energy density arising from tensor modes, the corresponding effective pressure, the effective equation of state, and the effective gravitational energy $\rho^{(T)}_\mathrm{eff}+3p^{(T)}_\mathrm{eff}$ contributing to the Raychaudhuri equation, all normalised to the background expansion. A positive $\Omega_\mathrm{eff}^{(T)}$ increases the effective Hubble rate. An effective equation of state $w^{(T)}_\mathrm{eff}<-1/3$ is equivalent to a positive $(\Delta R/R)^{(T)}$, and in this case the tensorial backreaction will act to accelerate the volume expansion. For an effective cosmological constant, $w^{(T)}_\mathrm{eff}=-1$ and $(\Delta R/R)^{(T)}=2\Omega^{(T)}_\mathrm{eff}$, while an effective dark energy with the equation of state $w_\mathrm{eff}=-8/9$ would produce $(\Delta R/R)^{(T)}=(5/3)\Omega^{(T)}_\mathrm{eff}$.

\section{Results}
\label{Sec:Results}
\subsection{General Considerations}
The expressions (\ref{TensorR}, \ref{TensorBacks}) are in general spatial integrations across real variables. The most frequent approach to these equations employs ergodic averaging (see for example \cite{Paranjape09,Clarkson09,Brown09-2} for recent discussions), working with domains large enough that a volume average tends towards the ensemble average. This approach has the advantage of being well-studied and amenable to both analytic and numerical work as the perturbations reduce to variables of the wavenumber only.

Using the Fourier convention
\be
  A(\mathbf{x})=\int A(\mathbf{k})e^{-i\mathbf{k}\cdot\mathbf{x}}\frac{d^3\mathbf{k}}{(2\pi)^3}, \quad
  A(\mathbf{k})=\int A(\mathbf{x})e^{i\mathbf{k}\cdot\mathbf{x}}d^3\mathbf{x}
\ee
the average of a quantity quadratic in linear perturbations can be rewritten as
\be
\label{AvA2}
  \av{A^2(\mathbf{x})}=a^3\int\dom{W}(\mathbf{x})A^2(\mathbf{x})d^3\mathbf{x}
  =a^3\iint\dom{W}^*(\mathbf{k})A(\mathbf{k}')A^*(\mathbf{k}'-\mathbf{k})\frac{d^3\mathbf{k}'}{(2\pi)^3}\frac{d^3\mathbf{k}}{(2\pi)^3}
\ee
where $W(\mathbf{x})$ is a window function specifying the domain $\mathcal{D}$, and
\bdm
  \dom{W}(\mathbf{x})=\frac{W(\mathbf{x})}{\Vd}, \quad
  \dom{W}(\mathbf{k})=\int\dom{W}(\mathbf{x})e^{-i\mathbf{k}\cdot\mathbf{x}}\frac{d^3\mathbf{k}}{(2\pi)^3} .
\edm
Taking an ensemble average of (\ref{AvA2}) and employing the definition of the power spectrum
\be
  \overline{A(\mathbf{k})A^*(\mathbf{k'})}=\frac{2\pi^2}{k^3}\left|A(k)\right|^2\mathcal{P}(k)\delta(\mathbf{k-k}')
\ee
we find that
\be
  \overline{\av{A^2(\mathbf{x})}}=\frac{R}{(2\pi)^3}\int\left|A(k)\right|^2\mathcal{P}(k)\frac{dk}{k} .
\ee
Here
\be
  R=\frac{a^3\int W(\mathbf{x})d^3\mathbf{x}}{\dom{V}}=\frac{\int W(\mathbf{x})\sqrt{h_0}d^3\mathbf{x}}{\int W(\mathbf{x})\sqrt{h}d^3\mathbf{x}}\approx 1
\ee
is the ratio between the domain volume evaluated on the background and evaluated on the perturbed surface. The results of \cite{Behrend08,Brown09,Paranjape08,Clarkson09} suggest that the averages of the scalar perturbations remain relatively small, implying that to a good approximation we can take $R\approx 1$. It is now straightforward to show that terms containing contractions of a tensor mode with a partial derivative, such as $\av{h_{ij}^T\pd^i\pd^j\psi}$, vanish upon ensemble averaging, since the tensor modes are transverse-traceless. It is worth emphasising that, since the tensor modes are intrinsically gauge-invariant, the vanishing couplings with scalar quantities with $R\approx 1$ now implies that our results will themselves be automatically gauge-invariant.

The tensor perturbation can be expanded into its two polarisation states by
\be
  h_{ij}^T=h_+A_{ij}^++h_\times A_{ij}^\times .
\ee
$A_{ij}^{+,\times}$ form an orthonormal basis and explicit forms could be found, for example, by aligning the coordinate system with the Fourier mode (see for example \cite{Landriau02,Brown06}). The two polarisation states are assumed to possess similar statistics, that is that
\be
  \overline{h_+(k)h^*_+(k')}\approx\overline{h_\times(k)h^*_\times(k')}\approx\overline{h(k)h^*(k)}
\ee
We assume the primordial gravitational waves to have been produced by some inflationary process, with a power spectrum
\be
  \mathcal{P}_T(k)=A_T\left(\frac{k}{k_*}\right)^{n_T}=A_T\left(\frac{k_*}{k}\right)^m .
\ee
$A_T$ is the amplitude of the primordial tensor perturbations, evaluated at a pivot wavenumber $k_*$, which is typically chosen to be $k_*\approx 0.05$Mpc$^{-1}$. In the simplest inflationary models, $A_T=rA_S$, where $A_S$ is the amplitude of primordial scalar perturbations. The tensor spectral index is $n_T$, which in simple models is approximately $n_T\approx -r/8$. The gravitational wave spectrum is therefore slightly red and we will generally employ the inverse spectral index, $m=-n_T$, which is a small, positive number. When a specific model is required we take $r=1/20$ and $m=-n_T=1/160$.

The tensor corrections to the Friedmann and Raychaudhuri equations (\ref{TensorR},\ref{TensorBacks}) can now be written as
\bea
\label{Corrections1}
  \Rd^{(T)}&=&-\frac{1}{2(2\pi)^3}\int k^2\mathcal{P}_T(k)\left|h(k)\right|^2\frac{dk}{k}=\beta\int_0^\infty k^2\left|h\right|^2\frac{dk}{k^{1+m}}, \\
  \Pd^{(T)}&=&-\frac{1}{(2\pi)^3}\hub\int\mathcal{P}_T(k)h(k)\dot{h}^*(k)\frac{dk}{k}=2\beta\hub\int_0^\infty h\dot{h}^*\frac{dk}{k^{1+m}}, \\
\label{Corrections2}
  \Qd^{(T)}&=&-\frac{1}{2(2\pi)^3}\int\mathcal{P}_T(k)\left|\dot{h}(k)\right|^2\frac{dk}{k}=\beta\int_0^\infty\left|\dot{h}\right|^2\frac{dk}{k^{1+m}}
\eea
with
\be
  \beta=-\frac{A_Tk_*^m}{2(2\pi)^3} .
\ee
At linear order, the kinematical backreaction is related to the energy carried by the gravitational waves, while the dynamical backreaction is related to their energy flux.

Gravitational waves evolve as
\be
\label{GravWaveEvol}
 \ddot{h}+2\hub\dot{h}+k^2h=\frac{\pi}{a^2}
\ee
where $\pi$ contains the fluid anisotropic stresses. In the following two sections we solve this equation and evaluate the backreaction terms (\ref{Corrections1}-\ref{Corrections2}) in radiation and matter domination respectively.

\subsection{Radiation Domination}
In radiation domination the photons and baryons are tightly coupled to one another and their anisotropic stresses driven towards zero. Neglecting the anisotropic stress of the neutrinos, we can set this term to zero. The gravitational wave in radiation domination is then
\be
  h=\frac{\sin(k\eta)}{k\eta}
\ee
where we have removed a decaying mode that otherwise diverges as $\eta\rightarrow 0$. Using the coordinate $x=k\eta$, the curvature correction and backreactions (\ref{Corrections1}-\ref{Corrections2}) become
\bea
 \Rd^{(T)}&=&\frac{\beta}{\eta^{2-m}}\int_0^\infty\sin^2x\frac{dx}{x^{1+m}}, \\
 \Pd^{(T)}&=&\frac{2\beta}{\eta^{2-m}}\int_0^\infty\left(\frac{\sin 2x}{2x}-\frac{\sin^2x}{x^2}\right)\frac{dx}{x^{1+m}}, \\
 \Qd^{(T)}&=&\frac{\beta}{\eta^{2-m}}\int_0^\infty\left(\cos^2x-\frac{\sin 2x}{x}+\frac{\sin^2x}{x^2}\right)\frac{dx}{x^{1+m}} .
\eea
For $m>0$ we can find closed-form exact solutions. The backreaction terms in radiation domination are
\bea
  \Rd^{(T)}&=&\frac{\beta}{\eta^{2-m}}\frac{\sqrt{\pi}}{2m}\frac{\Gamma\left(\frac{2-m}{2}\right)}{\Gamma\left(\frac{1+m}{2}\right)}, \\
  \Pd^{(T)}&=&-\frac{\beta}{\eta^{2-m}}\frac{2^m}{(1+m)(2+m)}\frac{\pi}{\Gamma(m)\sin(\pi m/2)}=-\frac{4m}{(1+m)(2+m)}\Rd^{(T)}, \nonumber \\
  \Qd^{(T)}&=&\frac{\beta}{\eta^{2-m}}\frac{2-m}{2^{2-m}(2+m)}\frac{\pi}{\Gamma(1+m)\sin(\pi m/2)}=\frac{2-m}{2+m}\Rd^{(T)} . \nonumber
\eea
In the limit $m\ll 1$ expected in a realistic model of inflation, these reduce to
\be
  \Rd^{(T)}\approx\frac{\beta}{2m\eta^{2-m}}, \quad \Pd^{(T)}\approx-2m\Rd^{(T)}, \quad \Qd^{(T)}\approx(1-m)\Rd^{(T)} .
\ee
In Papers I and II it was shown that the dynamical backreaction arising from scalar perturbations is an order of magnitude smaller than the scalar curvature, with the kinematical backreaction negligibly small. In contrast, the kinematic backreaction produced by inflationary gravitational waves in radiation domination is approximately equal to the scalar curvature while the dynamic backreaction is suppressed by one power of the spectral index.

The effective energy density (\ref{EnergyPressureE}) in radiation domination is given by
\be
  \Omega_\mathrm{eff}^{(T)}=-\frac{1}{6}\beta\eta^m\frac{2^m}{2+m}\frac{\pi}{\Gamma(1+m)\sin(\pi m/2)}\approx-\frac{1}{6m}\beta\eta^m
\ee
and the effective normalised pressure (\ref{EnergyPressureP}) is
\be
  \Pi_\mathrm{eff}^{(T)}=-\frac{2^m(1-4m+m^2)}{18}\beta\eta^m\frac{\pi}{\Gamma(3+m)\sin(\pi m/2)}\approx \frac{1}{3}\Omega_\mathrm{eff}^{(T)} .
\ee
Since $\beta<0$, both the effective energy density and pressure are positive definite. To linear order in $m$, the effective equation of state (\ref{wRay}a) is
\be
  w_\mathrm{eff}^{(T)}\approx\frac{1}{3}\left(1-5m\right),
\ee
demonstrating that in some respects the tensor backreaction fluid acts as an effective radiation. Finally, the normalised impact on the Raychaudhuri equation (\ref{wRay}b) is
\be
  \left(\frac{\Delta R}{R}\right)^{(T)}=\frac{2^m(2-3m-m^2)}{12}\beta\eta^m\frac{\pi}{\Gamma(3+m)\sin(\pi m/2)}
   =-\frac{2-3m-m^2}{2(1+m)}\Omega_\mathrm{eff}^{(T)} .
\ee
As $\beta<0$, $(\Delta R/R)^{(T)}<0$. The tensor backreaction in radiation domination then acts to decelerate the volume expansion. It is also worth noting that the effective fluid properties grow only slowly with time, $\Omega_\mathrm{eff}^{(T)}\propto\eta^m\propto a^m$ and so are almost constant.

\subsection{Matter Domination}
The situation in matter domination is more complicated. Continuing to assume a vanishing anisotropic stress is a good approximation while the photons remain tightly-coupled. During and after recombination, the photon energy density is low enough that, even though the photon anisotropic stress is non-vanishing, the approximation remains at least reasonable. The solution to (\ref{GravWaveEvol}) in a universe dominated by dust is given by
\be
\label{GeneralH}
  h=A_k\frac{k\eta\cos(k\eta)-\sin(k\eta)}{k^3\eta^3}+B_k\frac{k\eta\sin(k\eta)+\cos(k\eta)}{k^3\eta^3} .
\ee
The normalising functions $A_k$ and $B_k$ are found by matching the mode and its first derivative at equality, and in terms of $x=k\eta=kt\eta_e$ are given by
\be
  A_k=-2+\cos^2\left(\frac{x}{t}\right)-\frac{t}{x}\sin\left(\frac{2x}{t}\right), \quad
  B_k=\frac{x}{t}+\frac{1}{2}\sin\left(\frac{2x}{t}\right)+2\frac{t}{x}\sin^2\left(\frac{x}{t}\right) .
\ee
Here $t$ is the conformal time in units of the conformal time at matter/radiation equality, $t=1$ corresponding to the epoch of equality and $t\approx 110$ to the present epoch.\footnote{In deriving this value we have assumed $\Omega_m\approx 1$ and $H_0=70.1$kms$^{-1}$Mpc$^{-1}$.}  The form of these functions significantly complicates the full evaluation of the backreaction, but exact solutions can be found in a particular limiting case.

Taking $t\rightarrow\infty$, corresponding to the universe at very late times, yields closed-form solutions. In this limit, the functions in (\ref{GeneralH}) are heavily skewed towards a small $k\eta_e$, and we can expand the normalisation functions $A_k$ and $B_k$ to zeroth order,
\be
  A_k\approx -3, \quad B_k\approx 0 .
\ee
Solving (\ref{Corrections1}-\ref{Corrections2}) in this case gives the correction terms, for any $m>0$,
\bea
\label{MatterLabel1}
  \Rd^{(T)}&=&9\frac{\beta}{\eta^{2-m}}\frac{2^m}{(1+m)(2+m)(4+m)}\frac{\pi}{\Gamma(m)\sin(\pi m/2)}, \\
  \Pd^{(T)}&=&-72\frac{\beta}{\eta^{2-m}}\frac{2^m}{(1+m)(3+m)(4+m)(6+m)}\frac{\pi}{\Gamma(m)\sin(\pi m/2)} \nonumber \\ &=&-\frac{8(2+m)}{(3+m)(6+m)}\Rd^{(T)}, \\
\label{MatterLabel2}
  \Qd^{(T)}&=&9\frac{\beta}{\eta^{2-m}}\frac{(2-m)2^m}{(1+m)(2+m)(4+m)(6+m)}\frac{\pi}{\Gamma(m)\sin(\pi m/2)} \nonumber \\ &=&\frac{2-m}{6+m}\Rd^{(T)}.
\eea
For small $m$ these are
\be
  \Rd^{(T)}\approx\frac{\beta}{4\eta^{2-m}}, \quad
  \Pd^{(T)}\approx-\frac{8}{9}\Rd^{(T)}, \quad
  \Qd^{(T)}\approx\frac{1}{3}\Rd^{(T)} .
\ee
In contrast to both the situation in radiation domination and that of scalar perturbations at any time, all three corrections are of the same order of magnitude. Using (\ref{EnergyPressureE}), the effective energy density of the backreaction is now
\be
  \Omega_\mathrm{eff}^{(T)}=-3\beta\eta^m\frac{2^{m}}{(1+m)(2+m)(4+m)(6+m)}\frac{\pi}{\Gamma(m)\sin(\pi m/2)}\approx -\frac{1}{8}\beta\eta^m
\ee
which as in radiation domination is positive definite. The normalised effective pressure (\ref{EnergyPressureP}) is
\be
  \Pi_\mathrm{eff}^{(T)}=\beta\eta^m\frac{2^{1-m}\left(16+11m+m^2\right)}{(1+m)(2+m)(3+m)(4+m)(6+m)}\frac{\pi}{\Gamma(m)\sin(\pi m/2)}
    \approx-\frac{8}{9}\Omega^{(T)}_\mathrm{eff} .
\ee
To linear order, the effective equation of state (\ref{wRay}a) is
\be
  w_\mathrm{eff}^{(T)}\approx -\frac{8}{9}-\frac{17m}{54}.
\ee
The backreaction produced by gravitational waves acts to accelerate the universe with an equation of state close to that of a cosmological constant and intriguingly close to the current observational bounds \cite{Komatsu08}. Furthermore, since $m>0$, $w^{(T)}_\mathrm{eff}$ is slightly more negative than the zeroth-order approximation $w^{(T)}_\mathrm{eff}=-8/9$. The direct impact on the Raychaudhuri equation (\ref{wRay}b) is
\bea
  \left(\frac{\Delta R}{R}\right)^{(T)}&=&-3\beta\eta^m\frac{2^{1-m}\left(10+9m+m^2\right)}{(1+m)(2+m)(3+m)(4+m)(6+m)}\frac{\pi}{\Gamma(m)\sin(\pi m/2)} \nonumber \\
   &=&\frac{10+9m+m^2}{6+2m}\Omega_\mathrm{eff}^{(T)}
   \approx\left(\frac{5}{3}+\frac{17}{18}m\right)\Omega_\mathrm{eff}^{(T)} .
\eea
This is positive definite and larger than the effective energy density by approximately a third. As the effective equation of state suggested, the tensor backreaction deep in matter domination therefore acts as an effective dark energy. The evolution of the effective energy density and acceleration,
\be
  \Omega_\mathrm{eff}^{(T)}\propto\left(\frac{\Delta R}{R}\right)^{(T)}\propto\eta^m\propto a^{m/2} ,
\ee
is markedly slow. The emergence of this effective dark energy is entirely natural -- we have introduced no new physics and assumed only the existence of primordial gravitational waves with a red spectral index.

The different behaviour of the backreaction in the epochs of radiation and matter domination arises purely from the different nature of the expansion at these times -- and, in particular, how it modifies the dynamical backreaction. After matter/radiation equality the single mode $h\propto j_0(x)$ is converted into the two matter modes until, at very late times, the second matter mode is again negligible. Early in matter domination we therefore expect the backreaction to change significantly before settling down at late times in the limiting case found above.

\begin{center}
\FIGURE{
\includegraphics[width=0.49\textwidth]{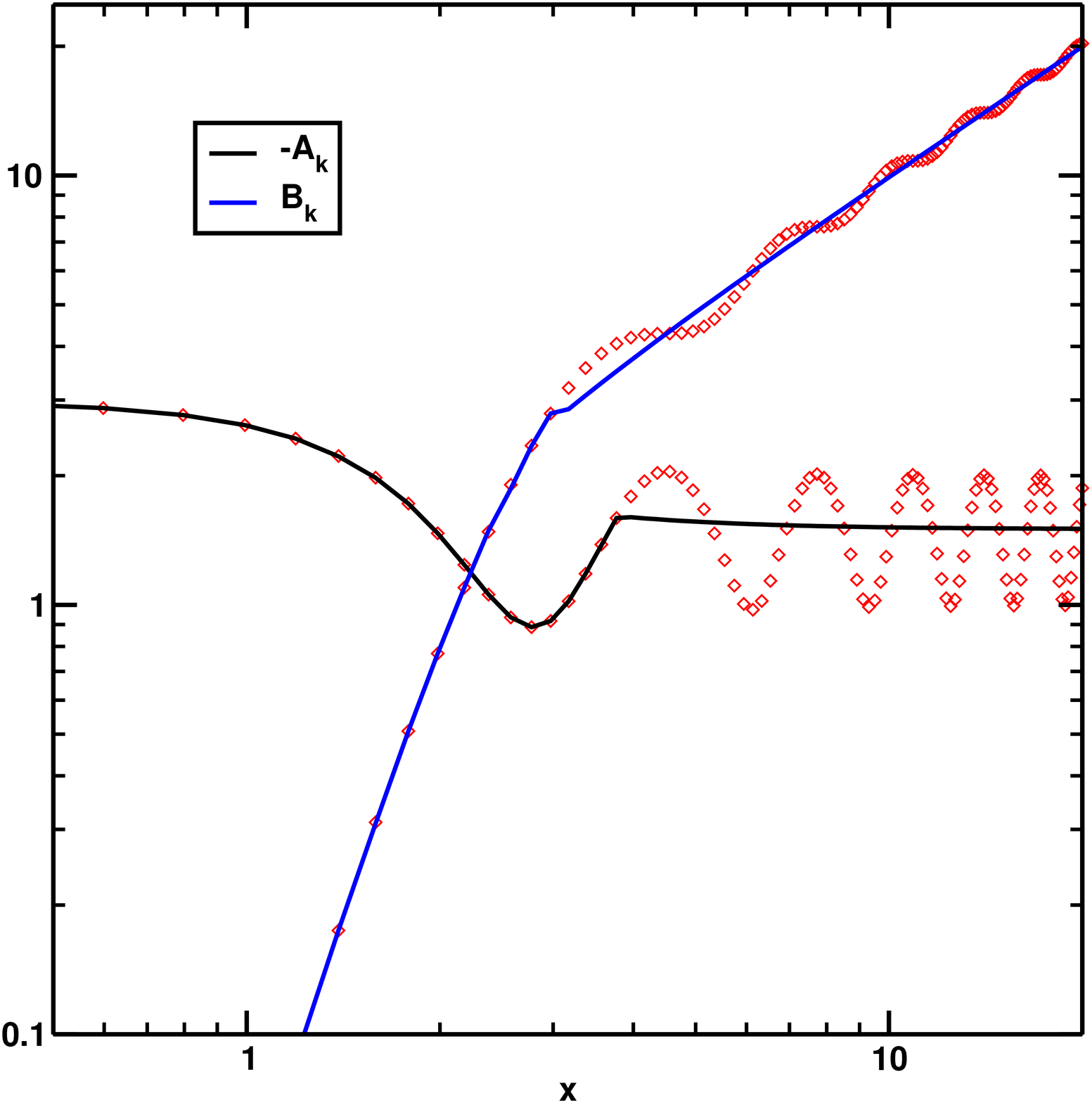}\;\includegraphics[width=0.49\textwidth]{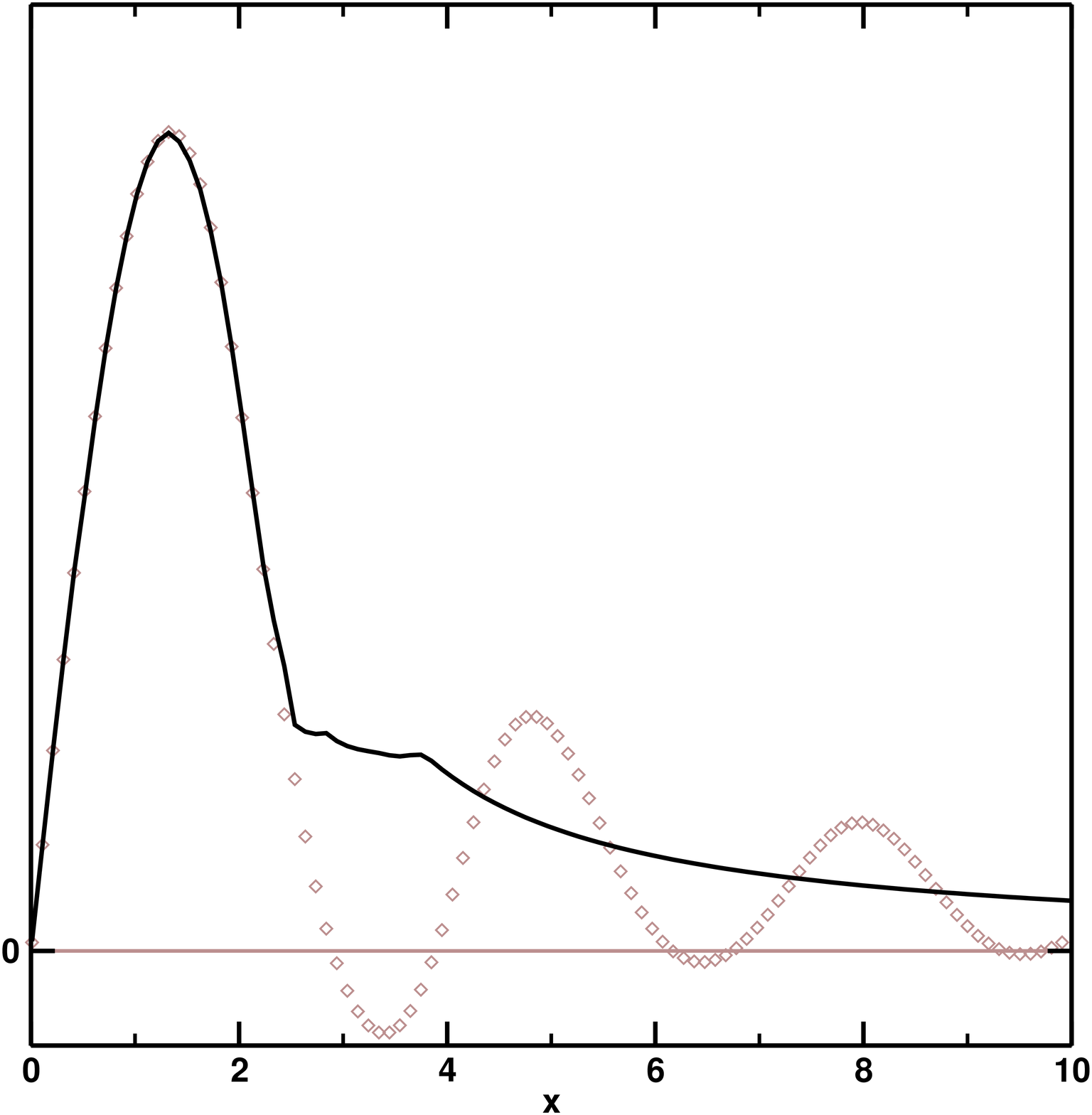}
\caption{The approximation scheme for $t=1$. Left: the normalisations $A_k$ and $B_k$ and their approximations. Right: the integrand of $\Rd$ and its approximation.}
\label{ApproxScheme}
}
\end{center}

The full integrals in matter domination are extremely complicated to solve. We employ an approximation scheme wherein we stitch together sequences of Taylor series for low values of $x$ with an averaged behaviour at higher $x$. For example, in $A_k$ and $B_k$ we Taylor-expand about $x\in\{1/2,1,3/2,2,5/2\}$, and employ the substitutions
\be
  \sin^2(x)\rightarrow\frac{1}{2}, \quad \cos^2(x)\rightarrow\frac{1}{2}, \quad \frac{2\sin(2x)}{x}\rightarrow\frac{\pi}{2x}
\ee
for $x\gtrsim 3$. This converts the intractable trigonometric functions into piecewise combinations of polynomials which are easily solved. Figure \ref{ApproxScheme} illustrates this approximation scheme for $A_k$ and $B_k$ and the integrand for $\Rd^{(T)}$.

We solve the resulting equations for the model with $m=1/160$ and the results are shown in Figure \ref{Fluid}. The left panel shows the evolution of the correction terms $\Rd$, $\Pd$ and $\Qd$ across matter domination. Shortly after equality, the curvature correction and kinematic backreaction decay rapidly, while the dynamic backreaction remains approximately constant. From a time of $\eta\approx 10\eta_e$ onwards (corresponding to a redshift of $z\approx z_e/1000\approx 100$) the decay slows and the corrections tend towards the late-time limiting cases (\ref{MatterLabel1}-\ref{MatterLabel2}). Also plotted in the figure for $t<1$ are the values evaluated for radiation domination. The backreaction terms over matter domination evolve smoothly between the two limiting cases, as expected. The right panel of figure \ref{Fluid} shows the evolution of the effective equation of state of the backreaction, which likewise tends smoothly from its radiation value of $w^{(T)}_\mathrm{eff}\approx 1/3$ to the present value of $w^{(T)}_\mathrm{eff}\approx -8/9$.
\begin{center}
\FIGURE{
\includegraphics[width=0.49\textwidth]{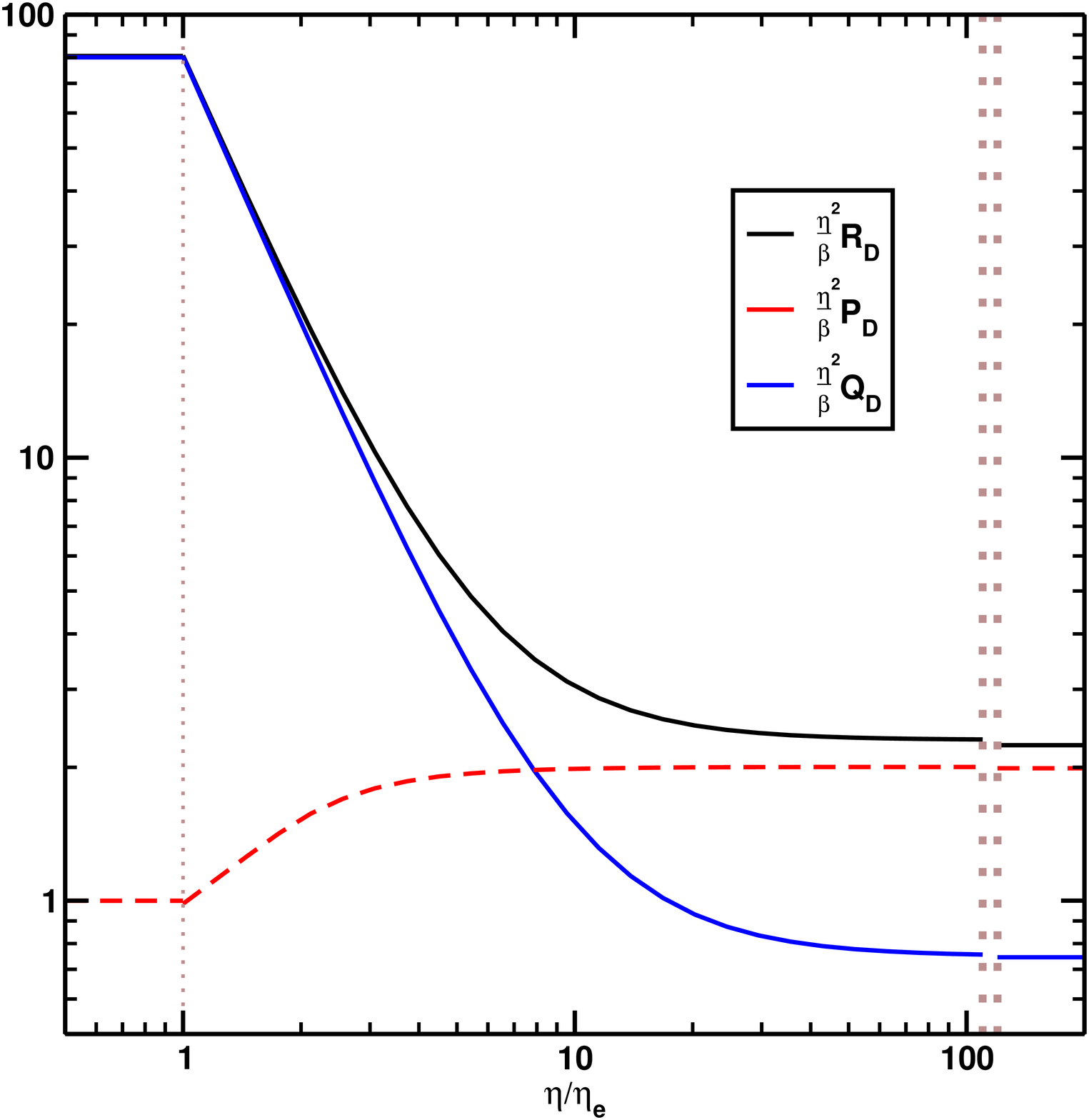}\;\includegraphics[width=0.49\textwidth]{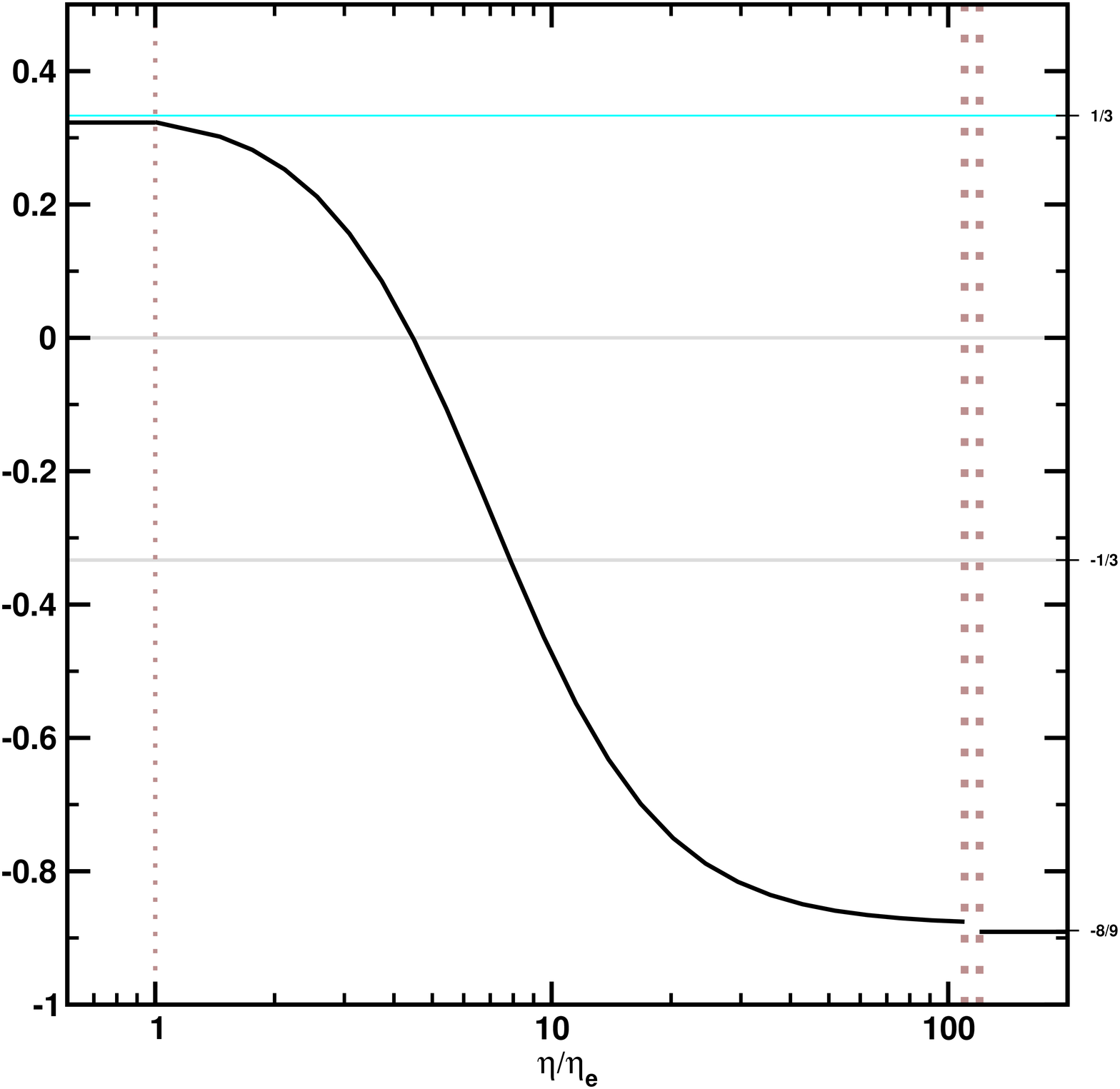}
\caption{The effective backreaction fluid as a function of $t=\eta/\eta_e$. Left: the corrections $\Rd^{(T)}$, $\Pd^{(T)}$ and $\Qd^{(T)}$, scaled by $\eta^2/\beta$. Right: the effective equation of state. For late times the backreaction acts as a dark energy.}
\label{Fluid}
}
\end{center}

Primordial gravitational waves, therefore, naturally give rise to a backreaction that acts in radiation domination as an additional radiative fluid but late in matter domination evolves to become a dark energy. The properties of this effective fluid are close to those required for a dark energy by observation. However, the effective energy density is unfortunately low. Deep in radiation and matter domination respectively it is
\be
  \Omega_\mathrm{eff}^{(T)}(\eta\lesssim\eta_e)=\frac{1}{6m}\frac{A_T}{(2\pi)^3}(k_*\eta)^m, \quad
  \Omega_\mathrm{eff}^{(T)}(\eta\gg\eta_e)=\frac{1}{16}\frac{A_T}{(2\pi)^3}(k_*\eta)^m .
\ee
If $A_T=rA_s$, with $r\approx 1/20$ and $A_S\approx 2.5\times 10^{-9}$, these are
\be
  \Omega_\mathrm{eff}^{(T)}(\eta\lesssim\eta_e)\approx 10^{-12}(k_*\eta)^m, \quad
  \Omega_\mathrm{eff}^{(T)}(\eta\gg\eta_e)\approx 10^{-14}(k_*\eta)^m .
\ee
Phrased differently, the backreaction from gravitational waves produces an error in the Hubble rate today of
\be
  \frac{\dom{H}}{H_0}\approx 10^{-7}(k_*\eta_0)^{m/2} .
\ee
Since $m\ll 1$ the normalisation $(k_*\eta)^{m/2}$ from the pivot wavenumber and conformal time will be of order unity. While the gravitational waves in such a universe certainly produce an acceleration that acts as a dark energy, it is unsurprisingly orders of magnitude below that which has been inferred from observation. Even so, the discovery of an accelerating volume-averaged expansion in a vanilla (FLRW+perturbations) system remains significant.

\section{Discussion}
\label{Sec:Discussion}
In this paper we have considered the backreaction that arises from primordial gravitational waves. While our motivation has primarily been those arising from inflation, the analysis is limited only by the requirement that the gravitational waves exhibit a red spectrum. Our first aim was to consider the tensor backreaction deep in radiation domination and compare it to that from the scalar modes. The two contributions are quantitatively similar: both the scalar and the tensor backreactions have effective energy densities which are approximately constant and effective equations of state $w_\mathrm{eff}\approx 1/3$, and the average spatial curvature $\Rd$ is the dominant contribution. However, while the scalar effective energy density is of order $\Omega_\mathrm{eff}\approx 10^{-8}-10^{-9}$, the tensor effective energy density is two orders of magnitude smaller. This suppression arises mainly from the small tensor/scalar ratio $r$. This implies that it is unlikely that the tensor backreaction has ever dominated over the scalars. Moreover, while in the scalar case the kinematical backreaction $\Qd$ is negligible and the dynamical backreaction $\Pd$ an order of magnitude smaller than $\Rd$, in the tensor case the kinematical backreaction is almost equal to the curvature correction while the dynamical backreaction is negligible.

Our second aim was to examine the changes that occur as the universe passes from radiation to matter domination and evolves towards the current epoch. The impact on the curvature correction and the kinematical backreaction is dramatic, both of them decaying rapidly in the matter era. The dynamical backreaction, in contrast, remains almost constant and indeed grows slightly. At a conformal time approximately ten times that of equality, the corrections tend towards asymptotic power laws $\sim\eta^{-(2+n_T)}$. The curvature correction $\Rd^{(T)}$ remains dominant, with the kinematical backreaction decaying only slightly faster. It is the dynamical backreaction that generates the difference between the radiation and matter cases: it tends towards a value approximately nine-tenths that of the curvature correction. The inhomogeneous corrections arising from tensor modes deep in matter domination, in direct contrast to those from scalars, are therefore all of equivalent magnitude. This then allows small changes in the corrections to produces differences in $w^{(T)}_\mathrm{eff}$ (\ref{wRay}a), with the ratios between them producing the intriguing nature of the late-time tensor backreaction.

The effective equation of state of the tensor backreaction tends for small $n_T$ towards $w^{(T)}_\mathrm{eff}\approx -8/9+17n_T/54$. It passes through $w_\mathrm{eff}^{(T)}=-1/3$ at a conformal time $\eta\approx 10\eta_e$, corresponding to a redshift $z\approx 100$ and approaches its asymptotic values at late times. The dark-energetic nature of the field thus emerges naturally throughout matter domination although it acts to accelerate the volume expansion from very early times. The effective energy density of the tensor backreaction is approximately constant and evolves with conformal time as $\Omega^{(T)}_\mathrm{eff}\sim\eta^{-n_T}\sim z^{n_T/2}$. It is, however, extremely low and suppressed from the backreaction in radiation domination by a factor of order $n_T$.

In a pure-dust universe, which would correspond to the universe in the far future (assuming $\Lambda=0$ and no other dark energy fluid), the backreaction reaches the asymptotic values. The tensor backreaction continues to slowly grow and exhibit an equation of state close to that of a viable dark energy. The effective fluid cannot be the observed dark energy for any sensible model since the amplitude is unsurprisingly low, but it nonetheless acts to accelerate the volume expansion. Given that we can find an effective dark energy from gravitational waves, it would be interesting to consider the more recent sources of gravitational radiation that emerge during the epoch of structure formation. It would be immediately interesting to examine the waves sourced by first-order scalar modes; in matter domination, first-order scalars act as a dark matter while first-order tensors act as a dark energy. How the second-order tensors act is an entirely open question, but their impact could be equivalent to the first-order tensors and they are of immediate significant interest.

We conclude that dark energy, with observationally reasonable equations of state, can arise from volume-averaging in pure relativity and in realistic systems, introducing no new physics beyond a slightly red distribution of primordial gravitational waves. While the energy scale of this ``dark energy'' is extremely low -- approximately fourteen orders of magnitude lower than the dark energy which is observed -- the significance remains: a natural dark energy can emerge in a vanilla FLRW+perturbations system.

\acknowledgments{IAB wishes to thank Christian Byrnes and Sami Nurmi for useful discussion and is supported by the Heidelberg Graduate School for Fundamental Physics.}

\bibliographystyle{JHEP}
\bibliography{Backreaction}

\end{document}